\documentclass[conference,compsoc,letterpaper]{IEEEtran}
\usepackage[hyphens]{url}

\usepackage{graphicx}
\usepackage{float}
\usepackage[caption = false]{subfig}
\usepackage{xcolor}

\newcommand{\eat}[1]{}

\begin{document}

\title{Online Password Guessability via Multi-Dimensional Rank Estimation}

\author{\IEEEauthorblockN{Liron David}
\IEEEauthorblockA{\textit{School of Electrical Engineering} \\
\textit{Tel Aviv University}\\
Ramat Aviv, 69978, Israel \\
lirondavid@gmail.com}
\and
\IEEEauthorblockN{Avishai Wool}
\IEEEauthorblockA{\textit{School of Electrical Engineering} \\
\textit{Tel Aviv University}\\
Ramat Aviv, 69978, Israel \\
yash@eng.tau.ac.il}

}

\maketitle
\thispagestyle{plain}
\pagestyle{plain}

\begin{abstract}
Human-chosen passwords are the a dominant form of authentication systems. Passwords strength estimators are used to help users avoid picking weak passwords by predicting how many attempts a password cracker would need until it finds a given password. 

In this paper we propose a novel password strength estimator, called PESrank, which accurately models the behavior of a powerful password cracker. PESrank calculates the rank of a given password in an optimal descending order of likelihood. 
PESrank estimates a given password's rank in fractions of a second---without actually enumerating the passwords---so it is practical for online use. It also has a training time that is drastically shorter than previous methods. Moreover, PESrank is efficiently {\em tweakable} to allow model personalization in fractions of a second, without the need to retrain the model; and it is {\em explainable}: it is able to provide information on {\em why} the password has its calculated rank, and gives the user insight on how to pick a better password.

Our idea is to cast the question of password rank estimation in a probabilistic framework used in side-channel cryptanalysis. We view each password as a point in a $d$-dimensional search space, and learn the probability distribution of each dimension separately. The dimensions represent the base word, plus a dimension for each possible transformation such as adding a suffix or using a capitalization pattern. Using this model, password strength estimation is analogous to side-channel rank estimation. 

We implemented PERrank in Python and conducted an extensive evaluation study of it. We also integrated it into the registration page of a course at our university. Even with a model based on 905 million passwords, the response time was well under 1 second, with up to a 1-bit accuracy margin between the upper bound and the lower bound on the rank. 
\end{abstract}

\section{Introduction}

\subsection{Background}
Text passwords are still the most popular authentication and are still in widespread use specially for online authentication on the Internet. Unfortunately, users often choose predictable and easy passwords, enabling password guessing attacks. Password strength estimators are used to help users avoid picking weak passwords. The most precise definition of password`s strength is \textit{the number of attempts that an attacker would need in order to guess it} \cite{dell2015monte}. 

A common way to evaluate the strength of a password is by heuristic methods, e.g., based on counts of lower- and uppercase characters, digits, and symbols (LUDS). Despite it being well-known that these do not accurately capture password strength \cite{weir2010testing}, they are still used in practice. 

Subsequently, more sophisticated, cracker-based, password strength estimators have been proposed. In a cracker-based estimator, either an actual password cracker is utilized to evaluate the password strength---or the estimator uses an accurate model of the number of attempts a particular cracker would use until reaching the given password. The main approaches have been based on, e.g., Markov models \cite{narayanan2005fast,ma2014study,durmuth2015omen}, probabilistic context-free grammars (PCFGs) \cite{weir2009password,komanduri2016modeling}, neural networks \cite{melicher2016fast,ur2017design} and others \cite{wheeler2016zxcvbn,guo2018lpse}.
Based on the well known phenomenon that people often use attributes of their personal information in their passwords (their names, email addresses, birthdays etc.),  \cite{li2017personal,houshmandusing,wang2016targeted} have proposed to tweak the prior models, based on personal information known about a given user's password. 

In this work we propose a novel addition to this line of research, called PESrank. Our goal is to provide a password strength estimator that enjoys the following properties: 
\begin{itemize}
\item It is a cracker-based estimator, that accurately models the behavior of a powerful password cracker. The modeled cracker calculates the rank of a given password in an optimal descending order of likelihood.
\item It is practical for {\em online} use, and is able to estimate a given password's rank in fractions of a second---i.e., without actually enumerating the passwords.
\item Has reasonable training time, drastically shorter than previous methods (some of which require days of training).
\item It is efficiently {\em tweakable} to allow model personalization, without the need to retrain the model.
\item It is {\em explainable}, and provides feedback on {\em why} the password has its calculated rank, giving the user insight on how to pick a better password.
\end{itemize}

\subsection{Contributions}

Our idea in the design of PESrank is to cast the question of password rank estimation in a probabilistic framework used in side-channel cryptanalysis. We view each password as a point in a $d$-dimensional search space, and learn the probability distribution of each dimension separately. This learning process is based on empirical password frequencies extracted from leaked password corpora, that are projected onto the $d$ dimensions. Once the $d$ probability distributions are learned, the a-priori probability of a given password is the {\em product} of the $d$ probabilities of its sub-passwords. 

Using this model, optimal-order password cracking is done by searching the space in decreasing order of a-priori password probability, which is analogous to side-channel key enumeration; likewise, password strength estimation is analogous to side-channel rank estimation. There is extensive research and well known algorithms for both problems in the side-channel cryptanalysis literature. We adopt a leading side-channel rank estimation (ESrank, \cite{dw19}) for use in PESrank---which accurately models an optimal key enumeration, or equivalently, password enumeration, algorithm. The ESrank algorithm also provides with accuracy guarantees with both upper-and lower-bounds on the true rank.

After a detailed evaluation of leaked password corpora we found that an effective choice is to use $d=5$ dimensions: base word, prefix, suffix, shift-pattern, and l33t transformation. 

The data structures which PESrank uses to represent its model are very simple: basically PESrank maintains a probability distribution for each of the $d=5$ dimensions. This implies that its training time is very reasonable, taking minutes-to-hours: We have been able to train PESrank on a corpus of 905 million leaked passwords \cite{1.4billion} within a few hours. This is in comparison with prior approaches that require days to train on corpora that are 10-20 times smaller. This representational simplicity also makes PESrank very efficient to tweak: we are able to personalize a general password model with per-user context extracted from, e.g., their email address, and do so in fractions of a second, without the need to retrain the model. 

The dimensions we picked are intuitive and follow common human practices in choosing passwords. And since PESrank maintains separate probability distributions for the $d=5$ dimensions, it is {\em explainable}. It is able to provide easy to understand feedback to the password owner, highlighting the popularity of her chosen base word, and the relative popularity of each of the other components. This is in contrast to prior methods, especially those based on neural networks, which are inherently unexplainable.

In order to demonstrate PESrank's capabilities as an online password strength estimator we implemented PERrank in Python and integrated it into the registration page of a course at our university. Even with a model based on 905 million passwords, the end-to-end response time in the browser was well under 1 second, with up to a 1-bit accuracy margin between the upper bound and the lower bound on the rank. 
This allowed us to run a proof-of-concept study on how students reacted to their passwords' strength estimates and to the explanations provided by PESrank: the anecdotal evidence we gathered is quite positive.

To test the power of PESrank tweakability we searched the corpus for evidence that people use parts of their usernames in their passwords. Not surprisingly, we discovered that 2.478\% of passwords' base words are equal to the username's base word, and 2.570\% use the same suffix. We also found strong evidence of password reuse: for \textbf{22\%} of the records in the corpus there exists at least one more record with the same \textit{name} and the same password. We then tweaked PESrank with this a-priori knowledge (per-user) to demonstrate the mount by which personal information 
can assist a password cracker.

We conducted an extensive evaluation study comparing PESrank's accuracy to prior approaches. 
In our study we used Ur et al.'s Password Guessability Service \cite{pgs} (PGS), which provides access to the Hashcat \cite{hashcat} and John the Ripper \cite{jtr} crackers, the Markov~\cite{narayanan2005fast} and PCFGs~\cite{weir2009password} methods, and to the neural-network method~\cite{melicher2016fast} (Monte-Carlo variant). We compared the ranks calculated by PESrank to the ranks obtained by these five password strength estimators. We show that PESrank (and, in fact, the optimal password cracker it models) is more powerful than previous methods: the model-based cracker can crack more passwords, with fewer attempts, than the password crackers we compared it to for crackable passwords whose rank is smaller than $10^{12}$.

{\bf Organization:} In the next section we introduce the background on side-channel key enumeration and rank estimation. In Section~\ref{sec:model} we describe our multi-dimensional password model. In Section~\ref{sec:dimensions} we describe the dimension selection process we applied. In Section~\ref{sec:performance} we describe the performance of PESrank  and the implementation details. In Section~\ref{sec:context} we discuss the benefit of using personal context when estimating the password strength. In Section~\ref{sec:usability} we present the usability of PESrank.  In Section~\ref{sec:comp} we compare PESank with existing methods.
In Section~\ref{sec:relatedWork} we discuss the related work and we conclude with Section~\ref{sec:concl}.

\section{Rank estimation and key enumeration in cryptographic side-Channel attacks} \label{sec:prelim}

Side-channel attacks (SCA) represent a serious threat to the security of cryptographic hardware products. As such, they reveal the secret key of a cryptosystem based on leakage information gained from physical implementation of the cryptosystem on different devices. Information provided by sources such as timing \cite{kocher1996timing}, power consumption \cite{kocher1999differential}, electromagnetic emulation \cite{quisquater2001electromagnetic}, electromagnetic radiation \cite{agrawal2003side,gandolfi2001electromagnetic} and other sources, can be exploited by SCA to break cryptosystems.

A security evaluation of a cryptographic device should determine whether an implementation is secure
against such an attack. To do so, the evaluator needs to determine how much time, what kind of computing power
and how much storage a malicious attacker would need to recover the
key given the side-channel leakages. The leakage of cryptographic implementations is highly device-specific, therefore the usual strategy for an evaluation laboratory is to launch a set of popular attacks, and to determine whether the adversary can break the implementation (i.e., recover the key) using ``reasonable`` efforts. 

Most of the attacks that have been published in the literature are based on a ``divide-and-conquer'' strategy. In the first ``divide'' part, the cryptanalyst recovers multi-dimensional information about different parts of the key, usually called subkeys (e.g., each of the $d=16$ AES key bytes can be a subkey). In the ``conquer'' part the cryptanalyst combines the information all together in an efficient way via key enumeration, for one of two purposes as follows.

\textbf{The Key Enumeration Problem}
The cryptanalyst obtains $d$ independent subkey spaces $k_1,...,k_d$, each of size $n$, and their corresponding probability distributions $P_{k_1},...,P_{k_d}$. The problem is to enumerate the full-key space in decreasing probability order, from the most likely key to the least, when the probability of a full key is defined as the product of its subkey's probabilities and test each full key in turn until the correct secret key is found.

A naive solution for key enumeration is to take the Cartesian product of the $d$ dimensions, and sort the $n^d$ full keys in decreasing order of probability. However this approach is generally infeasible due to both time and space complexity. Therefore several algorithms offering better time/space tradeoffs have been devised. The currently best optimal-order key enumeration is \cite{veyrat2012optimal}, with an $O(n^{d/2})$ space complexity, and near-optimal-order key enumeration algorithms with drastically lower space complexities are those of \cite{RSA,martin2015counting,bogdanov2015fast,poussiersimple,martin2018two}.

Unlike a cryptanalyst trying to extract the secret key, a security evaluator knows the secret key and aims to estimate the number of decryption attempts the attacker needs to do before he reaches the correct key, assuming the attacker uses the SCA's multi-dimensional probability distributions. Formally:

\textbf{The rank estimation problem:} Given $d$ independent subkey spaces of sizes $n_i$ for $i=1,\ldots,d$ with their corresponding probability distributions $P_{1},...,P_{d}$ such that $P_i$ is sorted in decreasing order of probabilities, and given a key $ k^*$ indexed by $(k_1,...,k_d)$, let $p^*=P_1(k_1) \cdot P_2(k_2) \cdot ... \cdot P_d(k_d)$ be the probability of $k^*$ to be the correct key. Estimate the number of full keys with probability higher than $p^*$, when the probability of a full key is defined as the product of its subkey's probabilities. 
In other words, the evaluator would like to estimate $k^*$'s \textit{rank}: the position of the key $k^*$ in the list of $n^d$ possible keys when the list is sorted in decreasing probability order, from the most likely key to the least. 
	
While enumerating the keys in the optimal SCA-predicted order is a correct strategy for the evaluator, it is limited by the computational power of the evaluator. Hence using algorithms to estimate the rank of a given key, without enumeration, is of great interest. Multiple rank estimation algorithms appear in the literature, the best of which are currently \cite{dw19,glowacz2015simpler,martin2015counting}. They all work in fractions of a second and generally offer sub 1-bit accuracy (so up to a multiplicative factor of 2).

\section{Multi-dimensional models for passwords} \label{sec:model}

\subsection{Overview}

The starting point in producing a password strength estimator is a leaked password corpus. The frequency of appearance of each leaked password provides an a-priori probability distribution over the leaked passwords. Given a hash of an unknown password, trying the leaked passwords in decreasing frequency order, is the optimal strategy for a password cracker---if the password at hand is in the corpus. To crack passwords that are not in the leaked corpus \emph{as-is}, password crackers rely on the observation that people often take a word, which we shall call the \emph{base word}, and mutate it using various transformations such as adding digits and symbols before or after the base word, capitalizing some of the base word's letters, or replacing letters by digits or symbols that are visually similar using ``l33t'' translations.

Our main idea is that if we can represent the list of base words as a dimension, and represent each possible class of transformations as another independent dimension, we can pose the password cracking problem as a key enumeration problem, and similarly, pose the password strength estimation as a rank estimation problem. Each dimension should have its own probability distribution: e.g., as we shall see, for the ``suffix'' dimension, the most probable in our training set is the ``empty'' suffix, with $\Pr=0.49$, followed by that of appending a digit `1' ($\Pr=0.062$) etc. Once we pose the password strength estimation question this way, we can use existing algorithms. A multi-dimensional password cracker would enumerate combinations of base word plus a transformation in every dimension, in decreasing order of the \emph{product} of per-dimension a-priori probabilities. For each combination it would apply the current set of transformations to the base word, and test the password. The matching multi-dimensional password strength estimator decomposes a given password into its base word and a transformation in every dimension, uses the model to calculate the a-priori probability of the password, and then estimates its rank.

Thus, we arrive at the following framework: First, identify meaningful classes of transformations, and find a suitable representation for each as a dimension. Next, build a probability distribution for each dimension using the training corpus, to create a model. Finally, use a good rank estimation algorithm with the model and evaluate its performance. 

\subsection{The Data corpus} \label{sec:corpus}

To study the statistical properties of passwords, and then to train our method, we used Jason's corpus of leaked passwords \cite{1.4billion}. This corpus contains 1.4 billion pairs of username and password, compiled from multiple leaked corpora: Yahoo, Target, Facebook, Hotmail, Twitter, MySpace, hacked PHPBB instances, and many other sources. We believe that Jason's corpus is a superset of the corpora used to train previous methods.

The ``username'' field in Jason's corpus is generally an email address, e.g., \texttt{adam1234@gmail.com}. It should be noted that the passwords provided in the corpus are not necessarily the passwords of the email accounts themselves (although they could be). Rather, the email addresses provided were harvested as the username for the breached entity in question, and the password provided is the password that was used with that username at the time of the breach.

After eliminating passwords that contain non-ASCII characters we obtained a corpus of 905,060,363 passwords. We then eliminated passwords of fewer than 8 characters, in order to make our training comparable to that of the methods accessible via PGS \cite{pgs}---in particular, the PGS implementation of the neural method \cite{melicher2016fast} is trained on passwords of length 8 or more. The result was a corpus of 905,047,301 passwords. 

From this corpus we sampled 300,000 username-password pairs, to serve as a test set. We split the test set into 10 separate samples, of 30,000 passwords each, and submitted all the sample sets to PGS for evaluation. 
Of the remaining passwords, we down-sampled the corpus into 5 different-size training sets, to explore the effect of the training-set size on the training time and the strength estimations. We used the following sizes for the smaller training sets: (1) 41 million username-passwords pairs (2) 113 million username-passwords pairs (3) 226 million username-passwords pairs (4) 452 million username-passwords pairs (5) and finally, 905 million username-passwords pairs.

\section{Selecting the dimensions} \label{sec:dimensions}

Our main goal in selecting the dimensions is generalization. For instance, if we recognize that a suffix of `1' has probability $p_1$, and that the base word `iloveyou' has probability $p_2$, a 2-dimensional model will \emph{implicitly} incorporate the password `iloveyou1', with probability $p_1\cdot p_2$, even if it does not appear as-is in the training corpus.

\subsection{The basic 3D model} \label{sec:3d}
In our simplest model
we divide each password into 3 sub-passwords: prefix, base word and suffix. The prefix consists of all the digits and symbols that appear to the left of the leftmost letter. The suffix consists of the digits and symbols that appear to the right of the rightmost letter. The base word is the string starting with the leftmost letter and ending with the rightmost letter. The base word can consist of letters, digits and/or symbols. For example if the password is the string `123abc45!' the prefix is `123', the base word is `abc' and the suffix is be `45!'. In case there are no letters in the password, the password itself is considered to be the base word, and the prefix and suffix are the empty strings. In case the password starts with a letter, the prefix is the empty string, and similarly, if the password ends with a letter, the suffix is the empty string.

\subsection{Using the model}

We use this basic 3D model description to demonstrate the process we followed in training and evaluating it. We followed the same process with the other, more sophisticated, models we describe in subsequent sections.

\subsubsection{The learning phase}
We first learn the distributions of the prefix, the base word and the suffix, using one of the training set sampled from the Jason corpus  \cite{1.4billion}---recall Section~\ref{sec:corpus}. Let these distributions be denoted by $P_1,P_2,P_3$ respectively.
For each password in the training set, we divide the password into its three sub-passwords, as described above, and increment the dimensional-frequency of each sub-password by 1. Finally we normalize the three lists of frequencies into probability distributions, and sort them in decreasing order. For instance, in the 3D model, using the 41 million password training set, the two most popular base words are `a' ($\Pr=0.0046$) and `password' ($\Pr=0.0035$), the two most popular prefixes are the empty string ($\Pr=0.910$) and `1' ($\Pr=0.0090$), and the most popular suffixes are also the empty string ($\Pr=0.497$) and `1' ($\Pr=0.062$).

\subsubsection{The estimation phase}
A model-based cracker based on, e.g., \cite{weir2009password} goes over the password candidates using an optimal-order enumeration. We can use a matching rank estimation algorithm such as \cite{dw19} to estimate the password guessability. Given a password $P$, we split into its sub-passwords $P=p^*||b^*||s^*$ where $p^*$ is a prefix, $b^*$ is a base word, and $s^*$ is a suffix. With this, using the three probability distributions $P_1, P_2, P_3$, we can apply a rank estimation algorithm such as \cite{dw19}. 
The algorithm estimates the number of 3-part passwords $p_i||b_j||s_k$ (split in the same way), whose probabilities obey
\[
P_1(i)\cdot P_2(j)\cdot P_3(k) \ge P_1(p^*)\cdot P_2(b^*)\cdot P_3(s^*).
\]
In other words, it estimates the number of guesses a model-based cracker would attempt before reaching the given password $P$.

If the password $P=p^*||b^*||s^*$ is not in the model, which means that at least one of the sub-passwords does not appear in the corresponding distribution, then we return -5 (following the behavior of PGS \cite{pgs} under analogous conditions).

\subsection{4D: Adding the shift pattern} \label{sec:4d}

The next level of sophistication is to add another dimension, to represent the ``shift pattern'': the pattern of upper-case and lower-case letters in the passwords. 

To begin with we split a password into its 3D sub-passwords as in Section~\ref{sec:3d}. We only consider the
shift pattern within the base word. Given a base word, it is clear which of its letters are shifted (upper case) and which are not. However, we have a choice how to represent the pattern. The obvious option is to represent the pattern as a binary string, with a `1' bit in position $j$ indicating that the $j$'th letter is capitalized. However, this representation is closely tied to the word's length: e.g., position 5 could indicate the last letter in a 5-letter word, or the before-last letter in a 6-letter word, etc. Our intuition, and a preliminary inspection of the Jason \cite{1.4billion} corpus, show that people tend to associate significance to the distance from the word's end: e.g., capitalizing the last letter is fairly common ($\Pr=0.0011$). Therefore we elected to represent the shift pattern as a list of positive and negative indices at which capital letters appear: The negative indices count from the word end, with -1 representing the rightmost letter. The positive indices count from the word start, with 0 representing the leftmost letter. To avoid ambiguity, both the negative and the positive indices do not exceed the middle index. 

We augment the 3D learning phase as follows. As in the 3D case, we divide each password into three sub-passwords: prefix, base word and suffix. Then, we find the indices at which there are upper case letters in base word, and represent their pattern using our positive/negative index representation. We increment the frequency of the pattern in the fourth dimension by 1.

Furthermore, before incrementing the base word's frequency, we ``unshift'' it, i.e., we ensure that all the base word letters are in lower case.  
If there is no letter in the base word, the password itself is the base word and the shift pattern is empty. Examples of the 4D decomposition appear in Table~\ref{tbl:4Ddecomposition}. The two most popular shift patterns we found in our training set are the empty transformation ($\Pr=0.923$) and `[0]', representing capitalizing the first letter ($\Pr=0.034$).

\begin{table}
\begin{center}
\footnotesize
 \begin{tabular}{||c|c|c|c|c||} 
 \hline
 password & prefix & base word & suffix & shift pattern \\ [0.5ex] 
 \hline\hline
 123PassworD & 123 & password & empty & [0,-1] \\ 
 \hline
 1234567890 & empty & 1234567890 & empty & empty \\ 
 \hline
 123qweASD & 123 & qweasd & empty & [-3,-2,-1] \\ 
 \hline
\end{tabular}
\end{center}
\caption{4D decomposition examples}
\label{tbl:4Ddecomposition}
\end{table}

Note that the shift-pattern dimension is not strictly independent of the base-word dimension: e.g., a shift pattern $t$ may refer to indices that are outside a short base word $b$, or $b$'s characters at the indexed positions may be symbols or digits (which do not have a capitalized form). In such cases the transformation $t$ degenerates into the null transformation. For a model-based password cracker, this dependence implies some inefficiency, since the cracker will test the same password multiple times, once for each shift-pattern that is equivalent to the null transformation for the current base word. The rank estimation accurately accounts for such a cracker's inefficiency. This means that a more sophisticated cracker can be developed: it could skip null transformations and save itself time. As we shall see in Section~\ref{sec:comp}, even though PESrank models a sub-optimal cracker, it actually under-estimates the ``ground-truth'' ranks fairly often---which means that such a model-based password cracker will perform better than the methods on which the ground truth is based upon.

\subsection{5D: l33t}

As observed by \cite{wheeler2016zxcvbn}, it is well documented that people sometimes mutate passwords using ``l33t'' transformations: replacing base word letters by digits and symbols that are visually similar. We elected to add the l33t transformation as a fifth dimension to our model. The l33t transformations we considered are shown in 
Table~\ref{tbl:leet}.

As in the 4D case (Section~\ref{sec:4d}) we need to devise a representation for the l33t pattern. In principle the l33t pattern depends on the position of the letter being mutated, and on the choice of replacement (Table~\ref{tbl:leet} shows that some letters have more than one l33t replacement). In this case we elected to ignore the positionality aspect. We numbered the possible l33t replacements from 1 to 14---e.g., transforming `a` into `4` is transformation 3---and represent the whole l33t transformation of a base word by a tuple of l33t replacement numbers. We assume that if a l33t replacement is applied then it is applied to \emph{all} the relevant letters in the base word. So following Table~\ref{tbl:leet}, the meaning we associate with a l33t pattern of `[1,3]' is ``replace all occurrences of o by 0 and all occurrences of a by 4''.

Note that our representation is unable to represent transformations in which the position matters: e.g., for a base word `aaaaaa' we cannot represent a l33t pattern that produces the password `aa44@@'.  

We make the following changes in the learning phase. As in the 4D case, we divide each password into four sub-passwords: prefix, base word, suffix and shift pattern. Then, we check which l33t replacements were applied to the base word using the options in Table~\ref{tbl:leet}. To prevent a collision between two different l33t symbols that represent the same letter, we track only the leftmost replacement (in the base word)  per row in Table~\ref{tbl:leet}. 
For each password we keep a tuple of all the l33t replacements that were done to the base word.

We increment the frequency of the tuple in the fifth dimension by 1. And as in the 4D model, before incrementing the the base word's frequency, we ``un-l33t'' it using the detected pattern.

If there are no l33t transformations in the base word, the base word remains as-is and the l33t pattern is empty.

For example: if the password is `g00dPa\$\$w0rD', the prefix is empty, the base word is `goodpassword', the suffix is empty, the shift pattern is `[4,-1]' and the l33t pattern is `[1,4]'.

\begin{table} 
\begin{center}
 \begin{tabular}{||c | c| c||} 
 \hline
 index & original letter & l33t \\ [0.5ex] 
 \hline\hline
 1 & o & 0  \\ 
 \hline
 2,3 & a & [@,4]  \\
 \hline
 4,5 & s & [$ \$ $,5]  \\
 \hline
 6 & e & 3  \\
 \hline
 7,8 & g & [6,9]  \\
 \hline
 9,10 &  t & [+,7]  \\
 \hline
 11 &  z & 2  \\
 \hline
 12,13 & i & [1,!]  \\
 \hline
 14 & x & $\%$  \\
 \hline
\end{tabular}
\end{center}
\caption{L33t transformations}
\label{tbl:leet}
\end{table}

Note that like the shift pattern in the 4D model (Section~\ref{sec:4d}), the l33t-pattern dimension is not independent of the base-word dimension: e.g., a l33t pattern $t$ may indicate a replacement of letters that do not appear in the base word, which means that for this word the pattern degenerates to another, simpler, l33t pattern, possibly the empty one. As before, this dependency introduces some inefficiency to a model-based password cracker.

The two most popular l33t patterns in our training set are the empty transformation ($\Pr=0.900$) and l33t pattern 3 `$4$' $\leftrightarrow$ `$a$' ($\Pr=0.007$). The total number of l33t patterns we detected in the corpus is 1596.

\subsection{Model enrichment} \label{sec:enrich}

\begin{table*}[t]
\footnotesize
\begin{center}
 \begin{tabular}{||c|c|c|c|c||} 
 \hline
  & 3D & 4D & 5D & 5D+ \\ 
 \hline\hline
 Total dimension lengths ($\cdot10^6$)  & 20.11 & 19.63 & 19.41 & 20.41 \\ 
 \hline
 Size (uncompressed, MB)  & 615 & 602 & 595 & 641 \\ 
 \hline
 Size (compressed, MB)  & 139 & 132 & 129 & 133 \\ 
 \hline
 Volume   & $1.73\cdot10^{19}$  & $7.73\cdot10^{23}$ & $1.21\cdot10^{27}$ & $1.28\cdot10^{27}$
 \\ 
 \hline
\end{tabular}
\end{center}
\caption{Model size using the 41 million passwords  training set}
\label{tbl:ModelSize}
\end{table*}

As we shall see, using a 5D model gives us good generalization capabilities beyond the training data. However, following \cite{li2014large, shay2010encountering, dell2010password}, we know that people have a tendency to choose passwords that contain dates and meaningful numbers. To take this observation into account, we enriched the probability distributions of the prefix, base word, and suffix dimensions, by adding strings that are not present in the training corpus. 
Every string that is added to a given dimension is added with a frequency $\epsilon =0.5$, to account for the fact that it didn't appear in the corpus. The resulting frequencies: those computed from the corpus (with values $\ge 1$), together with all enriched $\epsilon$ values, are then normalized into a probability distribution. We tested other values $\epsilon\ne 0.5$ but they all performed similarly as long as $\epsilon<1$. If the current enrichment string exists in the corpus with an empirical frequency $f$, we update its frequency to $f+\epsilon$. 
We enriched the model in two ways:
\begin{itemize}
    \item All the digit sequences of up to 4 digits were added to the prefix and suffix distributions. 
    \item All the digit sequences of length exactly 6 were added to the base word distribution.
\end{itemize}

Note that this enrichment gives good coverage of many date patterns: adding all the 6-digit strings to the base words covers all the dates that have patterns of ddmmyy or yymmdd or mmddyy, and most of the dates with a 4-digit year, such as 111998 (which could be either 1/Jan/1998 or Nov/1998). With the prefix and the suffix enrichments we cover all the 2-digit and 4-digit years such as 98 or 2003. An 8-digit date, e.g., with a ddmmyyyy format, as a stand-alone password, is also included in the model as both a 2-digit prefix plus a 6-digit base word, and as a 6-digit base word plus a 2-digit suffix. 

Now for a given password which is composed only of digits, the enriched model may include several options to reach this password by the model-based password cracker. As we saw with an 8-digit date example, a numeric password can be divided into prefix, base word, and suffix, in different ways, and the enriched model may include all the sub-passwords in the respective distributions. To account for this condition in the rank estimation, we added special handling of numeric passwords. For such a password, the PESrank algorithm iterates over all its possible divisions into 3 sub-passwords (of any length): for an $\ell$-digit password there are exactly $(\ell+1)(\ell+2)/2$ possibilities. For each division whose 3 sub-passwords appear in the model we calculate the password's probability. Finally, we return the rank of the division with the highest probability, since this is the division that will be encountered first by the optimal enumeration algorithm.

\subsection{Basic model evaluation}

In order to evaluate our 4 models we computed for each model: 
\begin{enumerate}

\item The percentage of passwords that would be cracked after a particular number of guesses: in other words, the Cumulative Distribution Function (CDF) of each model. More powerful guessing methods guess a higher percentage of passwords in our test set, and do so with fewer guesses: hence a better model has a CDF that rises more sharply and ultimately reaches a higher percentage.

    \item The model size: we calculated the Total Dimension Length---the sum of all the dimensions lengths i.e., $\sum_{i=1}^d n_i$. This metric counts the total number of (password-fragment, probability) pairs in the model. We also calculated the models' file sizes (uncompressed and compressed).
    
    \item volume: this is the total number of passwords that can be reached using our model. When the model has $d$ dimensions, whose sizes are $n_i$ for $i=1,\ldots,d$, then $\mbox{volume}=\prod_{i=1}^d n_i$.
\end{enumerate}

\begin{figure}
\includegraphics[width=\columnwidth]{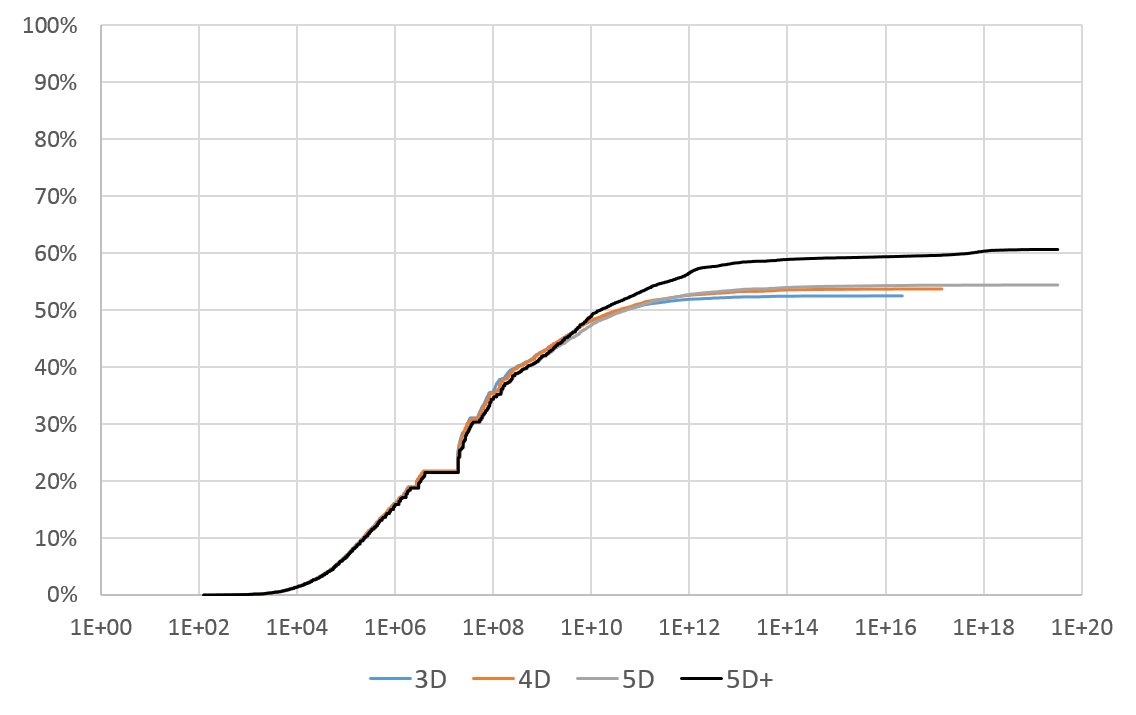}
\caption{Comparing the four models' performance on a 30,000 password test set, as a CDF: Each curve shows the fraction of passwords that can be cracked in up to $r$ attempts as a function of the rank $r$ (log scale).}
\label{figure:4modelcover}
\end{figure}

We trained the four models using the 41 million password training set, and evaluated them using 10 different test sets (recall Section~\ref{sec:corpus}). The results are shown in Figure~\ref{figure:4modelcover}. The 10 graphs of the different test sets look similar, therefore we show only one of them. 

In Figure~\ref{figure:4modelcover} the x-axis represents the number of guesses (log scale) and the y-axis shows the corresponding percentage of passwords in the test set that can be guessed in that number of guesses. As we can see, the method becomes more powerful with each added dimension, successfully cracking a greater fraction of the passwords, with fewer attempts.We also see that adding the enrichment by numeric strings as described in Section~\ref{sec:enrich} is very effective.

In Table~\ref{tbl:ModelSize} we can see the model sizes of the different models. The table shows that the model size \emph{decreases} with the Shift (4D)  and L33t (5D) dimensions: despite the fact that additional probability distributions are incorporated, the base word dimension shrinks due to different passwords collapsing into the same base word.

The enrichment step (5D+) does not grow the prefix and the suffix lists since after our training step, on the 41 million password training set, the prefix and the suffix lists already include all the 4-digits combinations. However, the enrichment step does increase the base word list by 1 million 6-digit combinations since we excluded passwords of fewer than 8 characters from training set (including the all-numeric 6-digit passwords in the corpus).

Even though the model size remains fairly stable with the added dimensions, 
Table~\ref{tbl:ModelSize} shows that the volume grows dramatically, from a raw corpus of about $10^8$ passwords to a password volume of $10^{19}$ in the 3D model, and by another 8 orders of magnitude reaching about $10^{27}$ with the 5D+ model.

We conclude that the enriched 5D+ model is superior to the simpler alternatives, and it's size is well within the capabilities of modern computers. In the remainder of this paper we use this enriched 5D+ model.

\subsection{Training time}

\begin{figure}[t]
\begin{center}
\includegraphics[width=\columnwidth]{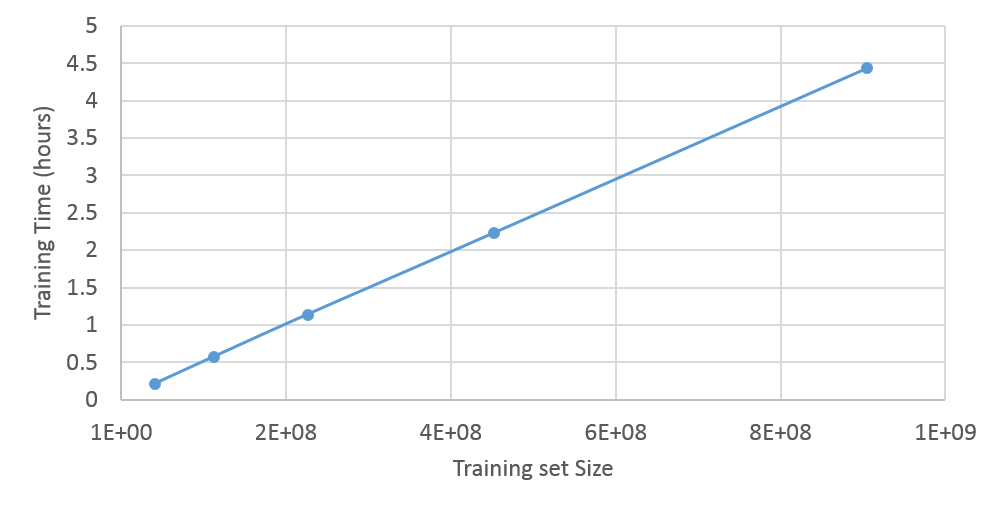}
\end{center}
\caption{Training time as function of the training set size.}
\label{figure:TrainingTime}
\end{figure}

We tested our Python implementation of PESrank's training on a 3.40GHz core 7 PC running Windows 8.1 64-bit with 32GB RAM. Figure~\ref{figure:TrainingTime} shows the time to train PESrank as a function of the training set size. The figure shows that the PESrank training phase is quite fast---much faster than reported for previous methods. It takes only 12 minutes to train PESrank on a corpus of 41 million passwords, in comparison to the days of training reported for the Markov \cite{ma2014study} or PCFG \cite{weir2009password} methods, that were trained on the similarly-sized ``PGS training set''. To train our method on 113 million passwords, it took only 34.5 minutes, in comparison to the days it took to train the neural method \cite{melicher2016fast} on the similarly-sized ``PGS++ training set'' (see more details in Section ~\ref{sec:perfoverview}). Because the PESrank training time is fast, we are able to train PESrank on a corpus of roughy 8 times the size of the ``PGS++ training set'', with 905 million passwords, and even on this corpus the training only took 4.5 hours.

\subsection{Storage requirements}
\begin{figure}[t]
\begin{center}
\includegraphics[width=\columnwidth]{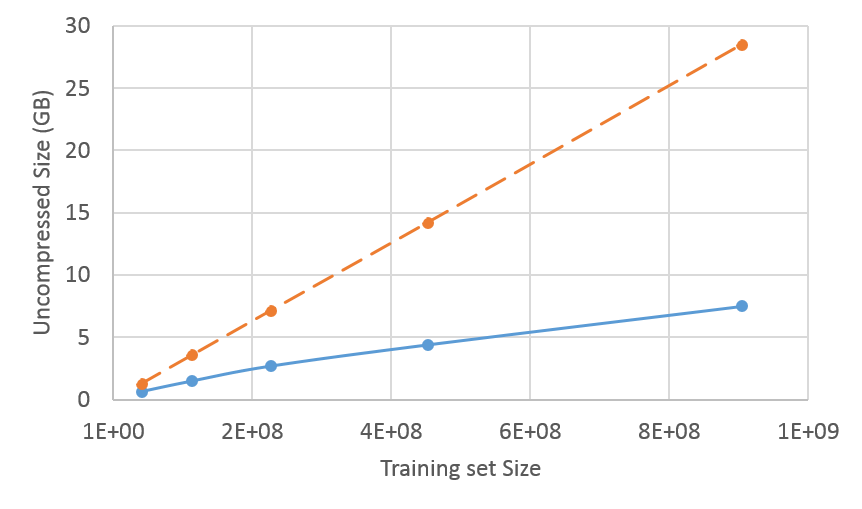}
\end{center}
\caption{The blue line is the uncompressed size of the 5D+enrichment model as function of the training set size. The dashed red line is the uncompressed training set size.}
\label{figure:TrainingSize}
\end{figure}

Beyond Table~\ref{tbl:ModelSize}, Figure~\ref{figure:TrainingSize} shows the effect of the training set size on the storage requirements. As we can see, the required storage grows more slowly than the training set size, reaching 7.5GB with the 905 million passwords training set: significantly less than the 40GB of the raw Jason corpus \cite{1.4billion}, and well within the capabilities of modern computers.

\section{PESrank performance and implementation details}
\label{sec:performance}

In the offline training step, PESrank goes over the training set passwords and creates five probability distribution lists stored in five files. Each file contains a list of $\{$value,probability$\}$ pairs. For the base word the value is the lowercase unl33t word. The representation of the other dimensions is as described in Section~\ref{sec:dimensions}.
The list in each file is sorted in increasing lexicographic order, to allow quick lookups. Note that the rank of the password fragment is not kept in the file - just its probability.

In the online test step, given a password, PESrank divides the password into its five dimension parts, then for each sub-password, it finds its probability using binary search in the corresponding probability distribution file. PESrank then multiplies the five probabilities to get the a-priori password probability. Finally, PESrank applies the rank estimation algorithm, ESrank \cite{dw19}, that receives the password probability and outputs upper and lower bounds for the rank. 
We also created an alternative implementation, as a serverless Google Cloud Function: in this implementation the probability distributions are placed in a MySQL database. The Google Cloud Function is meant to work as an API that can be embedded into any web site's password registration page.

The ESrank algorithm \cite{dw19} uses exponential sampling, which depends on a  tunable parameter $\gamma$ that affects the accuracy of the rank estimation and the time-and-memory trade-off. In \cite{dw19} the authors show that the ratio between the upper bound and the lower bound of a given password is bounded by $\gamma^{2d-2}$, therefore with $d=5$ dimensions we chose $\gamma$ to be 1.09 to achieve an estimation accuracy of less than 1 bit. 

\subsection{Reducing the online lookup time}

The PESrank lookup time, given a password, includes calculating the bounds of the password probability rank using the ESrank algorithm \cite{dw19}. The ESrank algorithm \cite{dw19} receives as input $d$ probability distributions lists and a password probability, uses exponential sampling to merge the $d$ lists into two, and then calculates upper and lower bounds of the rank of the given probability using only the last two merged lists. Since calculating these two merged lists only depends on the $d$ probability distribution lists, the merging step can be done in the offline training step. The online lookup step reads the last two lists and use them in calculating the rank of a given password. Thanks to the exponential sampling, the lengths of these two merged lists are small, (see  Table~\ref{tbl:performance}), therefore we placed them inline in the estimator code.

\subsection{Performance}

To train our method, we again used the Jason corpus \cite{1.4billion} using 905 million passwords to be our training set.

We implemented PESrank in Python. 
This code is publicity available at GitHub \cite{githubPESrank}.
We tested its performance using a test set of 30,000 passwords. The results are summarized in Table~\ref{tbl:performance}. The table shows that on average an estimation takes 33 msec, and under 1 sec in all cases, giving a good user experience.

\begin{table}[t]
\footnotesize
\begin{center}
 \begin{tabular}{||c|c||} 
 \hline
   Training time &  4.5 hours \\ 
 \hline
 Total space & 7.69GB  \\ 
 \hline
   Average estimation time per password &  0.033 sec \\ 
 \hline
   Maximum estimation time per password &  0.792 sec \\ 
 \hline
 Accuracy per password (upper/lower bound) & $\le$ 1 bit  \\ 
 \hline
 Combined length of the two merged lists & 884 integers \\ 
 \hline
\end{tabular}
\end{center}
\caption{PESrank performance when trained on 905 million passwords.}
\label{tbl:performance}
\end{table}

\section{Adding user personal information} \label{sec:context}

It is well known that it is advantageous to {\em customize} or tweak a general-purpose password model with a user`s personal information, if such information is available - e.g., email address, name, birthday, etc. \cite{castelluccia2013privacy,narayanan2005fast, li2017personal,houshmandusing,wang2016targeted}.

Conveniently, since PESrank uses a very simple model representation---essentially a sorted probability list of password elements per dimension---it is efficient to tweak, without the need to retrain.  

\subsection{Per-user context in the corpus} \label{sec:CheckContext}
Before evaluating PESrank ability to utilize personal information, we tested whether per-user context is present in passwords in the Jason corpus \cite{1.4billion}. The corpus comprises of \textit{pairs} of ``username''+password: recall Section~\ref{sec:corpus}.
However, the usernames in this corpus look like email addresses: ``name@domain.tld''. We used the `name' information and the `domain' as a source of personal information of the user.

For each username-password pair in the corpus, we divided the ``username'' into its name part (the text before the `@' in the username), which we denoted by \textit{name}, and its domain part (the text between the `@' and the `.' in the username), denoted by \textit{domain}. We then split the \textit{name} into its 3D dimensions, as described in Section~\ref{sec:model}: prefix, base word, and suffix. We split the password into its 3D dimensions as well. 
We then counted the number of passwords with:
\begin{enumerate}
    \item  prefix equal to the \textit{name} prefix.
    \item  base word equal to the \textit{name} base word.
    \item  suffix equal to the \textit{name} suffix.
    \item  base word equal to the \textit{domain}.
    \item  full password identical to the full username.
\end{enumerate} 

\begin{table} 
\footnotesize
\begin{center}
 \begin{tabular}{||c|c||} 
 \hline
 password prefix == \textit{name} prefix !=``''   & 0.033\%  \\
 \hline
 password base word == \textit{name} base word  &  2.478\%  \\ 
 \hline
 password suffix == \textit{name} suffix !=``''  &  2.570\%  \\
 \hline
 password base word == \textit{domain}   &  0.088\%   \\
 \hline
 password == username  &  0.062\%  \\ 
 \hline

\end{tabular}
\end{center}
\caption{Frequencies of context matches in the Jason corpus}
\label{tbl:contxtTable}
\end{table}

Table~\ref{tbl:contxtTable} provides the results. 
According to Table~\ref{tbl:contxtTable}, there is significant context information in the \textit{name}'s base word, and in the \textit{name}'s suffix: 2.478\% and 2.570\% respectively. The prefix dimension of the username seems to carry much less context, with less than 0.1\% of records in the corpus showing a match. 

We therefore conclude that the username in the email address carries significant contextual information about the password, specifically in its base word and suffix. 

Conversely, we find that ``Site context'' is weak in the Jason corpus: less that 0.1\% of the passwords matched the \textit{domain}. A possible reason is that the domain part of the email address is not a good indicator to the site at which the password was used.

Therefore, for the purposes of password cracking and password strength estimation, it is reasonable to assume that a-priori, the password suffix has a $2.570\%$ probability of matching the \textit{name}'s suffix, and the a-priori probability the password's base word matching \textit{name}'s base word is $2.478\%$.

\subsection{Using context in PESrank} \label{sec:usingContext}

Recall that PESrank uses a very simple model representation: it keeps a sorted probability list of sub-password per dimension. Therefore, all that is required to tweak one of its dimensions (e.g., adding a base word $w$ with a-priori probability $p$) is:
\begin{itemize}
    \item Let $p_0$ be the pre-tweaking probability of $w$ in the dimension's probability distribution ($p_0$ can be 0), and let $\Delta p = p-p_0$. Normalize the probabilities of all the other words in that dimension by multiplying each with $1-\Delta p$,
    \item Insert $w$ into its correct place in the sorted order, with probability $p$. If $w$ was already present in the distribution, update its probability to be $p$.
\end{itemize}
Note that in implementing PESRank we neither need to explicitly insert the new word $w$ into the sorted list nor to normalize all the other probabilities. 
The tweaking step can be done during the online lookup in constant time: given a sub-password $s$, if $s$ is equal to the personal information word $w$ then its corresponding probability would be $p$ (without inserting $w$ into the sorted list). If the given sub-password $s$ is different from $w$, then its corresponding probability is its original probability normalized by $1-\Delta p$ (we normalize only the queried probability).

To evaluate the advantages of the username context contribution we used 11,880 password-username pairs from our test set. First we calculated the strength of each password using the 5D+ enrichment model using the 41 million training corpus. Then, for each password, we tweaked the base word dimension with the \textit{name}'s base word at probability 2.478\%, tweaked the suffix dimension with the \textit{name}'s suffix at probability 2.570\%, and calculated the strength of the password using the tweaked model.

\begin{figure}[t]
\begin{center}
\includegraphics[width=\columnwidth]{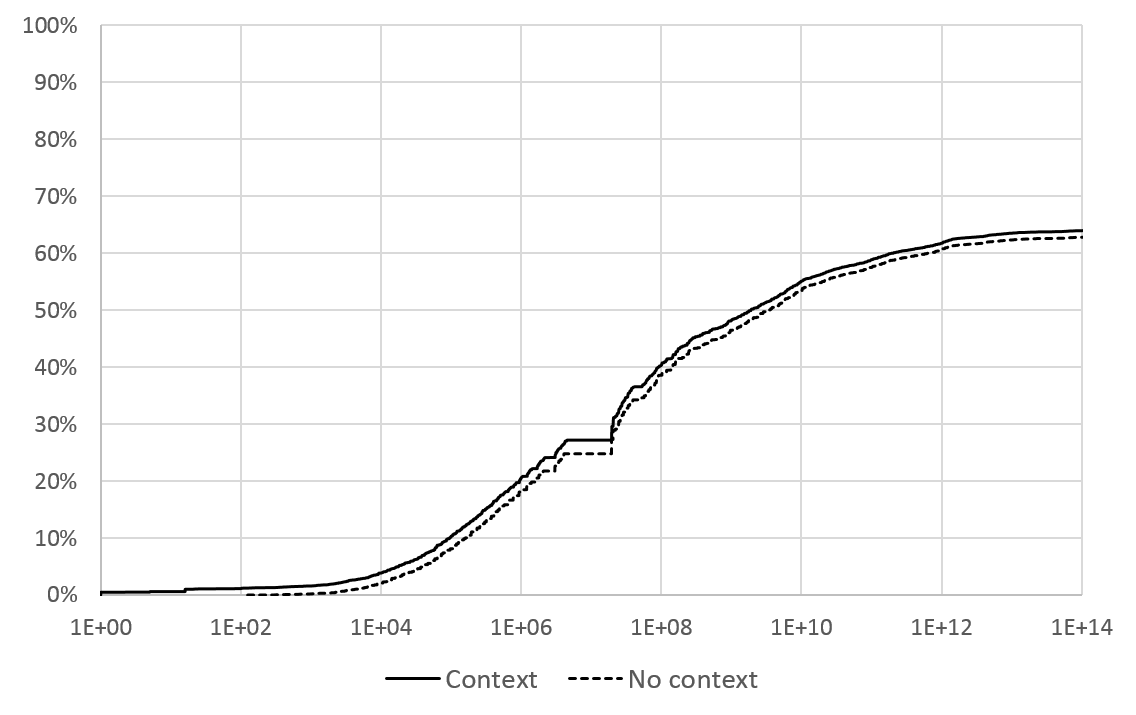}
\end{center}
\caption{The CDFs of PESrank with and without context, when trained on 41 million passwords.}
\label{figure:SiteContext}
\end{figure} 

Figure~\ref{figure:SiteContext} shows the results. In general we can see that adding context improves the estimation`s accuracy, and rank estimates are provided for 65.5\% of the passwords with context, 1.1\% more than without context. However, importantly, the improvement is most noticeable in the left side of the CDF curves: we can see that, with context, a significant number of passwords are discovered to be very weak: 20.6\% of the passwords can be cracked with fewer than $10^6$ attempts when using context, in comparison to 18.4\% without context. We conclude that incorporating even the very limited context present in the username can greatly benefit the password cracker, and deserves to be included in the strength estimate of a proposed password.

\subsection{Modeling password reuse} \label{sec:reuse}

It is well documented that people tend to reuse their previous passwords, either exactly, or with some variations. 

By a simple search of the Jason corpus \cite{1.4billion} we discovered that many users reuse their password \textit{exactly}. Specifically, as in Section~\ref{sec:CheckContext}, we extracted the \textit{name} part of each username, and we found that for \textbf{22\%} of the records in the corpus there exists at least one more record with the same \textit{name} and the same password. 

Therefore, when cracking, or evaluating the strength of a user's password, if we know previous passwords belonging to a user with the same \textit{name}, it is reasonable to assume that one of her known passwords or their variants is being reused with an a-priori probability of 22\%. Note that a properly configured site would not retain a user`s old passwords but only their hashes. However if the username exists in a leaked password corpus like Jason, then historical passwords become available.

We used this information to tweak PESrank in the following way:
For each user in the test set:
\begin{enumerate}
    \item Let $L$ denote the list of all the passwords (whose length is greater than 8 and contain only ASCII) in the Jason corpus \cite{1.4billion} that belong to usernames with the same \textit{name}, excluding the current password.
    \item Let $f_i$ be the frequency of each unique password in $L$.
    \item Split each password in $L$ into its 3D dimensions.
    \item Tweak the context of each prefix/base word/suffix dimension so its corresponding probability will be 0.22$\cdot f_i$ (as in Section~\ref{sec:usingContext}), and normalize the distributions.
\end{enumerate}

\begin{figure}[t]
\begin{center}
\includegraphics[width=\columnwidth]{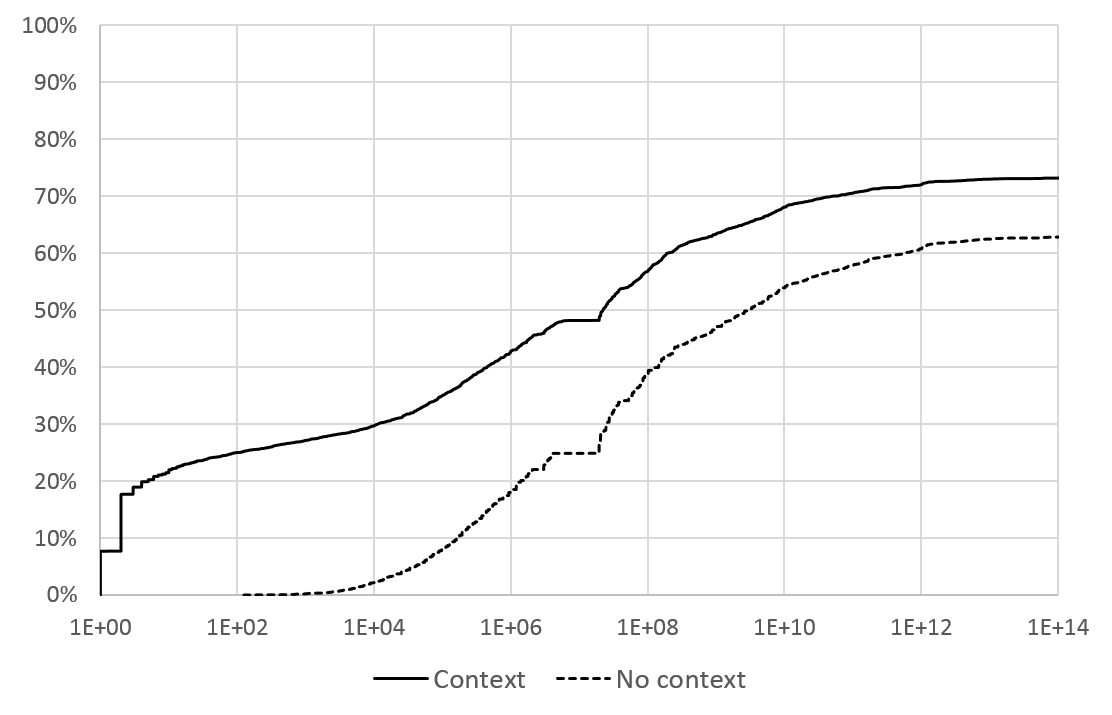}
\end{center}
\caption{The CDFs of PESrank with and without password-reuse context, when trained on 41 million passwords. }
\label{figure:PasswordContext}
\end{figure}

To evaluate the information in historical passwords, we sampled 5,000 username-password pairs from our test set, and evaluated their strength with the context-aware PESrank using both the username context of Section~\ref{sec:usingContext} and the password-reuse context (if such context was available), when our model is trained on 41 million passwords.
Figure~\ref{figure:PasswordContext} compares its CDF to that of the 5D+ model (as in Figure~\ref{figure:4modelcover}).
The figure clearly shows that when using both username and password-reuse context, there is a dramatic improvement in the cracking power (and a dramatic decrease in password strength). Again, the most important improvement is on the left side of the CDF: 42.8\% of the passwords can be cracked with fewer than $10^6$ attempts when using password-reuse context.

We draw several conclusions from this experiment. First, we re-affirm that knowing the user's previous passwords provides the password cracker with very valuable information. Second, that a good password strength estimator should use such history as context, to accurately model the cracker's capability---and that resources such as Jason's corpus \cite{1.4billion} already provide a wealth of historical data on millions of users. And third, that the PESrank algorithm is well suited to use such context since it is easily tweakable without time consuming training.

\section{Usability of PESrank}
\label{sec:usability}

\subsection{A proof of concept study}

\begin{figure}[t]
\begin{center}
\includegraphics[width=\columnwidth]{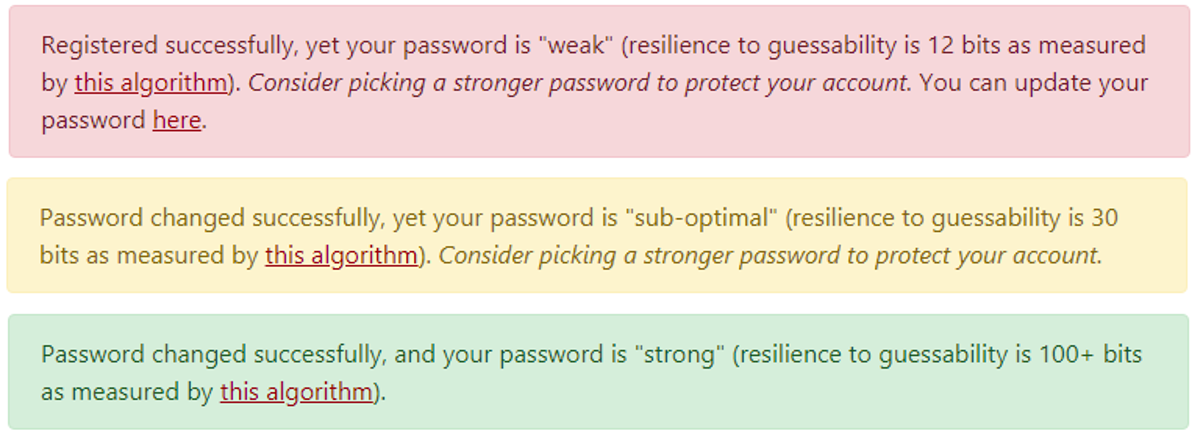}
\end{center}
\caption{The possible messages shown by the registration page.}
\label{figure:message}
\end{figure}

We integrated PESrank into the registration page of the Infosec course at our university. The system provides users with a gentle ``nudge'': it accepts weak passwords, yet tells the owners they are weak, and makes it easy for them to try again. The system outputs three different messages, see Figure~\ref{figure:message}: passwords with strength below 30 bits are considered `weak' (red), strengths between 30--50 bits are considered `sub-optimal' (yellow) and strengths above 50 bits are considered `strong' (green). 

During the registration we only saved the password strength and \emph{not the password itself} for statistical analysis, as approved by university's ethics review board.

The total time from clicking on the Register button until the browser shows the feedback message (including password registration, strength estimation, network delays, and browser rendering) is well under 1 second. The increase in registration time due to the strength estimation was negligible and qualitatively unnoticeable.

\begin{figure}[t]
\begin{center}
\includegraphics[width=\columnwidth]{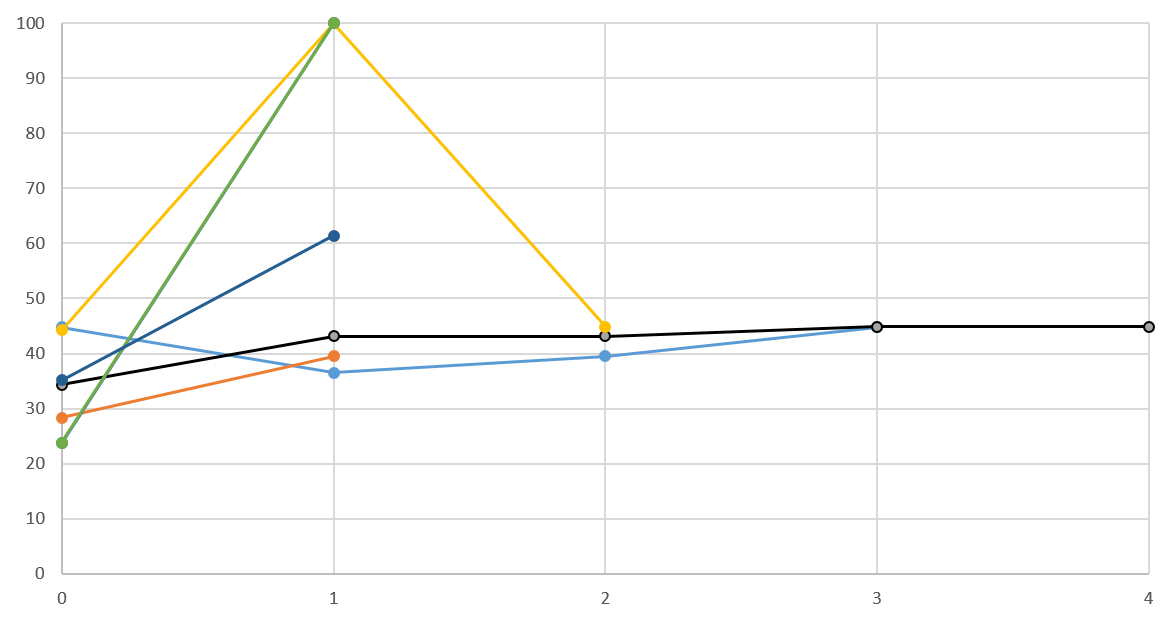}
\end{center}
\caption{The password strength versus password-change number for the 7 students who changed their password: index 0 indicates the strength of the initial password chosen by each student.}
\label{figure:change}
\end{figure}

There were 98 students who registered to this course: The median password strength of the first password chosen by the students was 41.51 bits, with the weakest having strength of 14.14 bits. Out of the 98 student only 7 students changed their passwords to stronger passwords. The median strength of these students' first passwords was 34.32 bits, and the median strength of their final passwords was 44.88 bits: a significant improvement.

In Figure~\ref{figure:change}, we can see the evolution of passwords strengths of the seven students who changed their password (there are two students whose lines are overlap due to similar strength choices). The figure shows that 5 students indeed picked a stronger password in their first change---one of whom later changed the password a second time in favor of a weaker password. Interestingly, two students changed their passwords 3 and 4 times, respectively, without significantly improving their strength.

\begin{figure*}[t]
\begin{center}
\includegraphics[width=2\columnwidth]{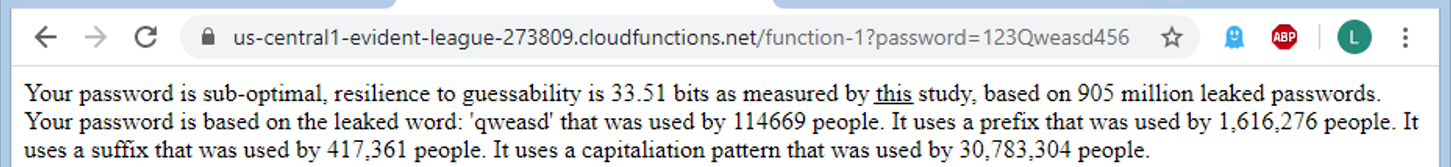}
\end{center}
\caption{PESrank using Google Cloud Function}
\label{fig:Google}
\end{figure*}

\subsection{Explainability} 

The anecdotal evidence from the proof of concept leads us to realize that just providing the password strength is not enough: a good strength estimator should give the user guidance on how to pick a better password.

One of the advantages of PESrank is that it is inherently very ``explainable''. As part of its calculation it discovers the a-priorty probability (and frequency) of each sub-password - and this information can be shown to the user. E.g., in the latest version of the code, for the password NewY0rk123 we provide the following feedback:
``Your password is sub-optimal, its guessability strength is 32 bits, based on 905 million leaked passwords. Your password is based on the leaked word: 'newyork' that was used by 129,023 people. It uses a suffix that was used by 17,631,940 people. It uses a capitalization pattern that was used by 592,568 people. It uses a l33t pattern that was used by 4,395,598 people''---See Figure~\ref{fig:Google} for a screenshot of the Google Cloud Function implementation.

This tells the user that (a) picking an unleaked base word is crucial, that (b) they use a very common suffix and that (c) a simple l33t transformation is only marginally effective. 
And most importantly - it teaches that the split into the 5 dimensions is something password crackers know about and take advantage of. We are planning to conduct a wider scale experiment in the future using the improved code.

\section{Comparison with existing methods} \label{sec:comp}

\subsection{Overview}\label{sec:perfoverview}

In order to test the power and accuracy of PESrank, we compared it to  multiple prier approaches. We first compared its guessing power, and its storage requirements, to those of four cracker-based methods offered in PGS \cite{pgs}: 
(1) the Markov model \cite{ma2014study}; 
(2) the PCFGs model \cite{weir2009password} with Komanduri`s improvements \cite{komanduri2016modeling};  (3) Hashcat \cite{hashcat}; (4) John the Ripper \cite{jtr} mangled dictionary models. Then we compared it to the neural network-based model of Melicher et al. (with Monte-Carlo estimation)~\cite{melicher2016fast}. 
Finally we tested the accuracy of PESrank as an estimator, versus the heuristic estimator, zxcvbn \cite{wheeler2012zxcvbn,wheeler2016zxcvbn}.
In all cases we used the algorithms' default settings and training data. 

Since we used the implementations of Neural, PCFG, Markov, hashcat and JtR, as made publicly available through the PGS service, we had no control over the corpora the PGS administrators used to train the algorithms---nor do we have the precise set of corpora that they used. According to the PGS site, they trained the PCFG, Markov, hashcat and JtR algorithms on 6 corpora,  totalling 33 million passwords, plus 6 million natural language words, collectively called the ``PGS training set''. The Neural algorithm was trained on passwords from 26 corpora, totalling 105 million passwords plus 6 million natural language words, called the ``PGS++ training set''. As far as we can tell the Jason \cite{1.4billion} corpus includes all the passwords that are in the PGS++ training set.

Therefore, to have a fair comparison, we trained PESrank on two training sets: one containing 41 million passwords, for comparison with the
PCFG, Markov, hashcat and JtR algorithms, and one containing 113 million passwords for comparison with the Neural algorithm. We also trained PESrank method on larger training sets (226 million, 452 million and 905 million passwords) in order to see how the training set size affects its performance.

\subsection{Comparison to cracker-based \& neural methods}

\begin{figure}[t]
\begin{center}
\includegraphics[width=\columnwidth]{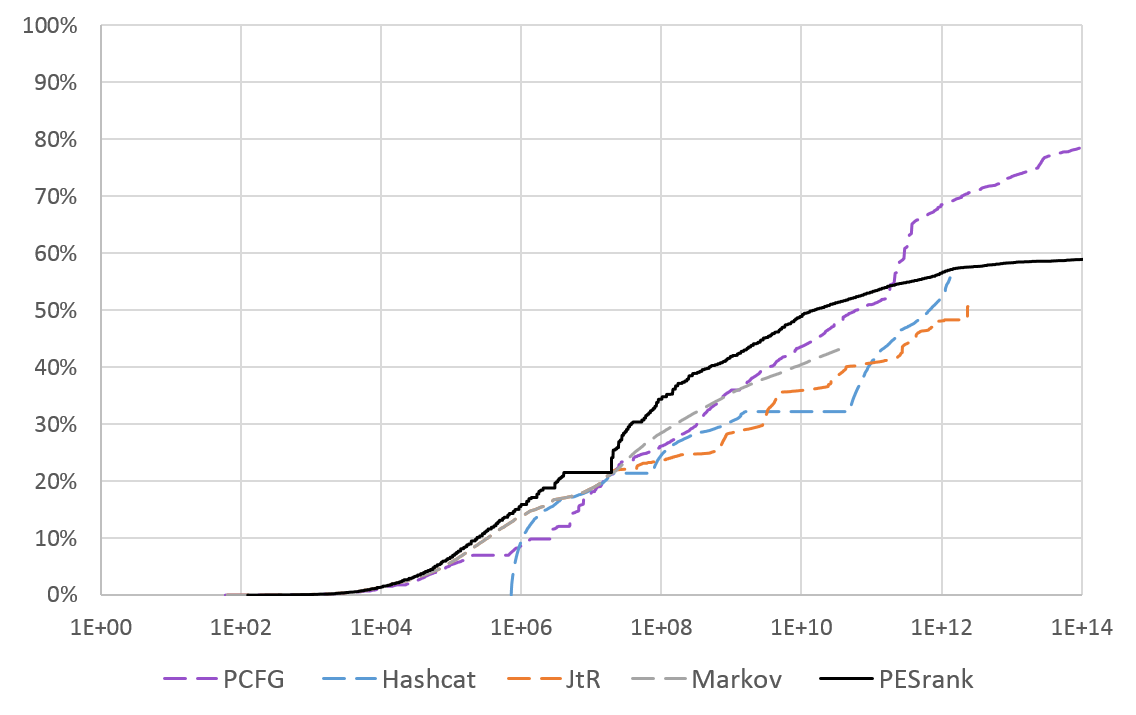}
\end{center}
\caption{The CDFs of PESrank versus PCFG, Markov, Hashcat and JtR, when PESrank is trained on 41 million passwords and the other methods are trained on the PGS training set. }
\label{figure:versusPGS40}
\end{figure}

\begin{figure}[t]
\begin{center}
\includegraphics[width=\columnwidth]{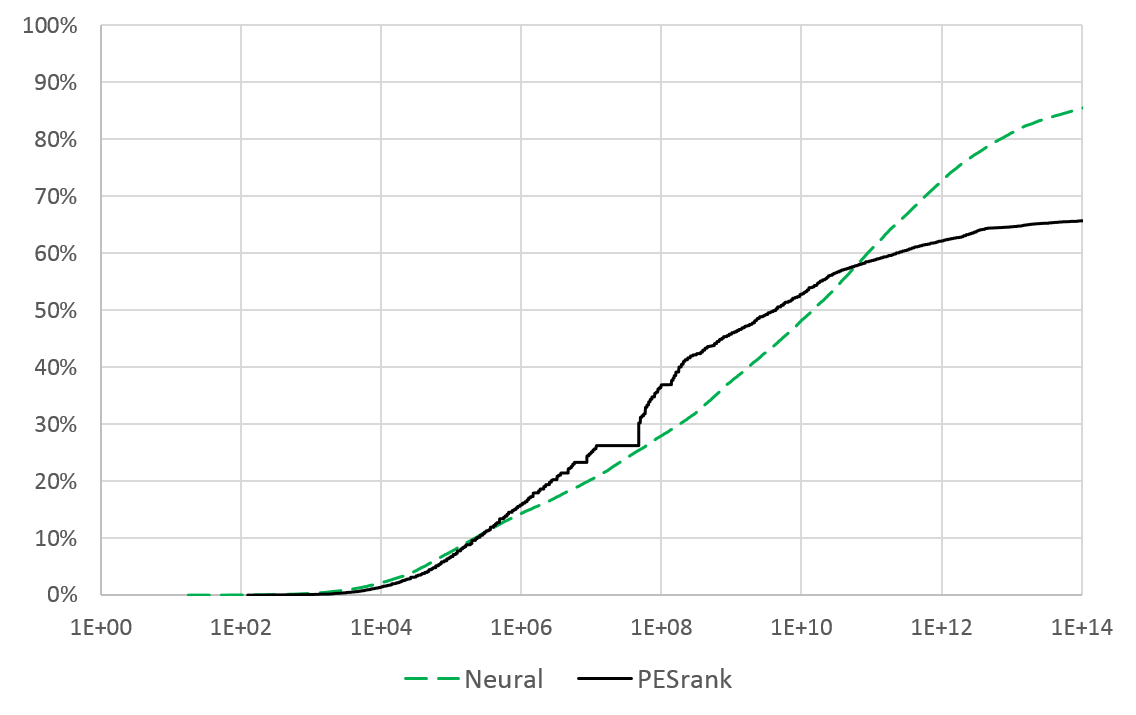}
\end{center}
\caption{The CDFs of PESrank versus Neural, when PESrank is trained on 113 million passwords and Neural is trained on the PGS++ training set.}
\label{figure:versusPGS110}
\end{figure} 

\begin{figure}[t]
\begin{center}
\includegraphics[width=\columnwidth]{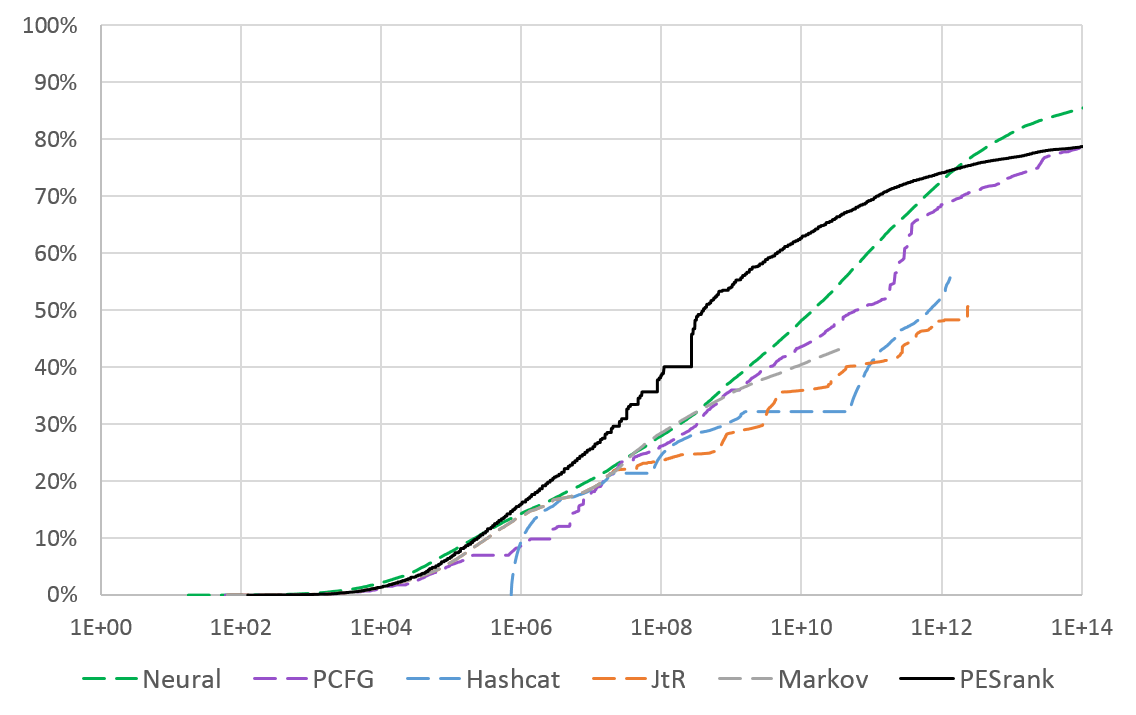}
\end{center}
\caption{The CDFs of all the method, when each method is trained on all the passwords available to it: PESrank trained on the 905 million password corpus, Neural trained on the PGS++ training set, and the rest trained on the PGS training set. }
\label{figure:versusPGSall}
\end{figure}

Figure~\ref{figure:versusPGS40} shows the CDFs comparing PESrank, trained on 41 million passwords with PCFG, Markov, hashcat and JtR algorithms, that are trained on the similarly sized PGS training set.
The figure shows that PESrank outperforms the other approaches for ``practically crackable'' passwords: Up to ranks around $10^{11}$, PESrank shows that our model based cracker is able to crack a higher percentage of passwords in the test set, and do it with fewer attempts than the other methods. While PCFG is able to provide lower ranks for passwords beyond rank $10^{11}$, we argue that it is more valuable to the user to learn that her password is crackable, versus just how uncrackable it is (e.g. for a PCFG rank of $10^{14}$).

Figure~\ref{figure:versusPGS110} shows the results comparing PESrank, trained on 113 million passwords, with the Neural method (with Monte Carlo estimation) which is trained on a similar corpus size (PGS++). Again, we see that PESrank outperforms the Neural method for ``practically crackable'' passwords, while Neural is able to better estimate the strengths of the very strong passwords.

Note that the Neural algorithm is apparently able to provide a rank for every possible password, and its CDF always reaches 100\%: we tested it on 20-character passwords, including letters, digits and symbols, generated uniformly at random by the ``Random.pw'' \cite{random.pw}. On such passwords Neural returned strength estimations of about $10^{41} \approx 2^{120}$, which is a fair estimate assuming 6-bits of entropy per password character. Thus Neural will always surpass PESrank on the right-hand side of the CDF, since PESrank will never return a rank that exceeds the password volume in the model (recall Table~\ref{tbl:ModelSize}).

Figure~\ref{figure:versusPGSall} shows the CDFs of all the methods we compared, each trained on the maximal training set available to it. When PESrank is trained on a 905 million password training set, its advantage over the other methods, as provided by the PGS service \cite{pgs}, grows, and only Neural surpasses PESrank, for ranks beyond $10^{12}$. While this figure mostly demonstrates the advantage of using a larger training set, it also shows that PESrank is actually able to digest such a large training set, due to its fast training time, whereas the other methods' ability to do so in reasonable time is currently unknown.

\subsection{Storage requirements}
Table~\ref{tbl:diskSpace} summarizes the storage space of the different methods, as reported by \cite{melicher2016fast}---where, unlike in the PGS service \cite{pgs}, the authors trained the earlier methods on the PGS++ training set. For comparison we provide the PESrank storage for the training set of size 113 million passwords. The table shows that the Neural network requires the lowest amount of storage (60MB) on the server-side, while PESrank requires a larger, yet very reasonable 1.4GB storage, and significantly less than PCFG.
\begin{table}[t]
\footnotesize
\begin{center}
 \begin{tabular}{||c|c|c|c|c|c||} 
 \hline
  PCFG & Markov & Hashcat & JtR & Neural &  PESrank \\ 
 \hline\hline
 4.7GB & 1.1GB  & 756MB & 756MB & 60MB &  1.4GB \\ 
 \hline
\end{tabular}
\end{center}
\caption{Storage requirements for the various methods as reported by \cite{melicher2016fast} when all methods are trained on the PGS++ corpus and PESrank is trained on 113 million passwords.}
\label{tbl:diskSpace}
\end{table}

\subsection{Comparison to a heuristic estimator}

We also compared the accuracy of PESrank with that of the zxcvbn \cite{wheeler2016zxcvbn,wheeler2012zxcvbn} heuristic.
We used the open source Python code of zxcvbn \cite{wheeler2016zxcvbn}. 

As a ground truth for this comparison, for each password, we used the minimum number of attempts needed to guess a password by running PGS \cite{pgs}, in its \textsc{min auto} configuration mode. In this mode PGS returns the minimum number of guesses among (1) the Markov model\cite{ma2014study}; (2) the PCFG model \cite{weir2009password} with Komanduri`s improvements \cite{komanduri2016modeling}; (3) Hashcat \cite{hashcat}; (4) John the Ripper \cite{jtr} mangled dictionary models; and (5) Neural Networks \cite{melicher2016fast} (with Monte-Carlo estimation).

\subsubsection{Estimator quality metrics} \label{sec:metrics}

Following \cite{wheeler2016zxcvbn}, given a test set $S$, on each sampled password $x_i \in S$ we measure an algorithm`s estimation error by computing its order-of magnitude difference $\Delta_i$ the from the PGS ground-truth:
\begin{equation}
    \Delta_i=\log_{10}\frac{g_{alg}(x_i)}{g_{pgs}(x_i)}
\end{equation}
where $g_{alg}$ is the rank estimate of the algorithm and $g_{pgs}$ is the minimum rank of the five PGS methods. Note that $\Delta_i$ can be positive (if the algorithm over-estimates the ground truth) or negative if it under-estimates.

We also define the following three metrics: \emph{over-estimation}, \emph{accurate} and \emph{under-estimation}, which give a sense of how frequently an estimated rank is significantly above or below the ground-truth and how frequently it is close to the ground truth. Specifically, \emph{over-estimation} is the fraction of passwords for which the estimated rank is two orders of magnitude ($\times 100$) more than ground-truth, and similarly for \emph{under-estimation}. \emph{accurate} is the fraction of passwords for which the estimated rank is within two orders of magnitude from the ground truth. Using the definition of $\Delta_i$ we obtain the following expressions:
\begin{equation}
     \mbox{\emph{over-estimation}}=\frac{1}{|S|}|\{x_i \in S : \Delta_i>2\}|
\end{equation}

\begin{equation}
     \mbox{\emph{accurate}}=\frac{1}{|S|}|\{x_i \in S : -2\le \Delta_i \le 2\}|
\end{equation}

\begin{equation}
     \mbox{\emph{under-estimation}}=\frac{1}{|S|}|\{x_i \in S : \Delta_i<-2\}|
\end{equation}

\subsubsection{Accuracy comparison} \label{sec:accuracy}

\begin{figure}[t]
\begin{center}
\includegraphics[width=\columnwidth]{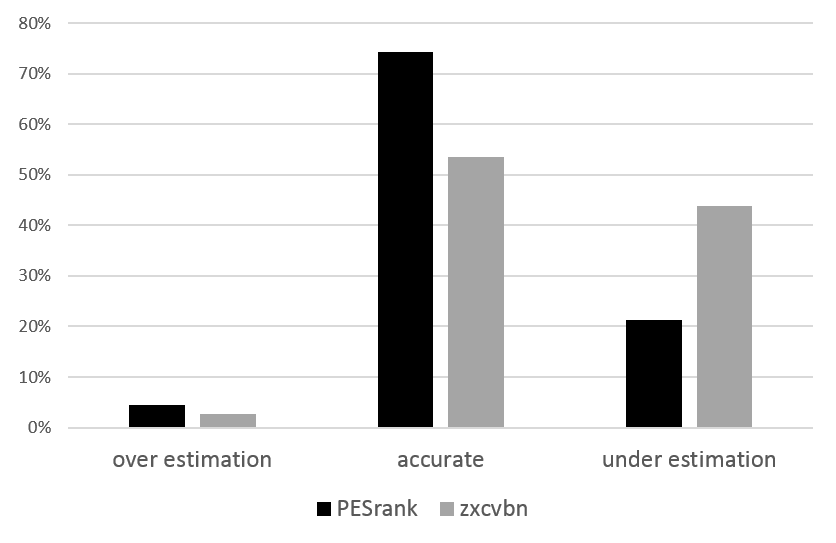}
\end{center}
\caption{The fractions of over-estimation, accurate, and under-estimation of PESrank and zxcvbn with a test set of 30,000 passwords.}
\label{figure:over}
\end{figure}

In Figure~\ref{figure:over} we see the over- and under- estimation of PESrank and zxcvbn using the metrics defined in Section~\ref{sec:metrics}. The figure shows that the PESrank method is more accurate than the zxcvbn: our model has the highest total number of passwords correlated with the ground truth. 

Moreover, the figure shows that zxcvbn under-estimates $43.87\%$ of the passwords. 
In contrast, PESrank is more aligned with the ground-truth, only under-estimating $21.21\%$ of them. 
Notice that PESrank under-estimation is actually advantageous, since PESrank is backed by a model-based password cracking algorithm that can achieve its predictions. It means that such a cracker can do better than the ``ground-truth''.

\section{Related work}
\label{sec:relatedWork}

Password strength measurement often takes one of two conceptual forms: heuristic pure-estimator approaches, and cracker-based approaches. In the latter, either an actual password cracker is utilized to evaluate the password strength---or the meter uses an accurate model of the number of attempts a particular cracker would use until reaching the given password. In heuristic pure-estimator approaches, the model provides a strength estimate directly from attributes of the password, without an accompanying passwords cracker that can achieve the predicted number of attempts. PESrank belongs to the cracker-based approaches.

\subsection{Heuristic pure-estimator approaches}

The earliest and probably the most popular methods of password strength estimation are based on LUDS: counts of lower- and uppercase letters, digits and symbols. The de-facto standard for this type of method is the NIST 800-63 standard \cite{burr2004electronic,grassi2017nist}. It proposes to measure password strength in entropy bits, 
on the basis of some simple rules such as the length of the password and the type of characters used (e.g., lower-case, upper-case, or digits).
These methods are known to be quite inaccurate \cite{de2014very}.

In 2012 Wheeler proposed an advanced 
password strength estimator 
\cite{wheeler2012zxcvbn}, that extends the LUDS approach by including dictionaries, considering l33t speak transformations, keyboard walks, and more. Due to its easy-to-integrate design, it is deployed on many websites. The meter's accuracy was later backed up by scientific analysis \cite{wheeler2016zxcvbn}. 

In 2018 Guo et al.~\cite{guo2018lpse} proposed a lightweight client-side meter. It is based on cosine-length and password-edit distance similarity. It transforms a password into a LUDS vector and compares it to a standardized strong-password vector using the aforementioned similarity measures. 

Such pure-estimator approaches have the advantage of very fast estimation---typically in fractions of a second---which makes them suitable for online client-side implementation. However, they do not directly model adversarial guessing so their accuracy requires evaluation.

\subsection{Cracker-based approaches}

\begin{table*}[h]
\begin{center}
 \begin{tabular}{||c|c|c|c|c||} 
 \hline
 &  PCFG & Neural & Neural+MonteCarlo & PESrank \\ 
 \hline\hline
 Training Time &  hours/days & hours/days & hours/days &  12 minutes \\ 
 \hline
 Lookup Time & offline  & offline  &  online &  online \\ 
 \hline
 Tweak Time & $\le$ 1 second  & hours/days  & hours/days & $\le$ 1 second  \\ 
 \hline
 Storage Requirement & highest  & lowest & lowest & medium  \\ 
 \hline
 Explainability & Maybe  & No & No & Yes  \\ 
 \hline
 Accuracy  &  exact & exact & unknown & up to 1 bit  \\ 
 \hline
\end{tabular}
\end{center}
\caption{Comparison to existing methods, when each method is trained on 41 million leaked passwords.}
\label{tbl:relatedWork}
\end{table*}

Software tools are commonly used to generate password guesses \cite{goodin2013anatomy}. The most popular tools transform a wordlist using mangling rules, or transformations intended to model common behaviors in how humans craft passwords. Two popular tools of this type are Hashcat \cite{hashcat} and John the Ripper \cite{jtr}.

These tools typically run until a timeout is triggered. If they crack the given password before the timeout then their accuracy is perfect: they can report exactly how many attempts they used. However, if they fail to crack the password by the timeout, they do not estimate how many more attempts would have been necessary. Since they generally take a long time to run (minutes to hours, depending on the timeout setting) their usefulness as online strength estimators is limited.

A probabilistic cracker method, based on a Markov model, was first proposed in 2005 \cite{narayanan2005fast}, and studied more comprehensively subsequently \cite{castelluccia2012adaptive,ma2014study,durmuth2015omen}. Conceptually, Markov models predict the probability of the next character in a password based on the previous characters, or context characters.
This method calculates the rank of a given password by enumerating over all possible passwords in descending order of likelihood. This enumeration is computationally intensive and limits the guess numbers of the methods, therefore the Markov model is impractical as an online strength estimator. In addition, in order to personalize this method for each user (i.e., giving a better estimation based on each user personal information), this method should be retrained for each user separately \cite{castelluccia2013privacy}. Since the training takes days, this method cannot be personalized for different users in real-time.

In 2009 Weir et al.~\cite{weir2009password} proposed a method which uses probabilistic context-free grammars (PCFGs). The intuition behind PCFGs is that passwords are built with template structures (e.g., 6 letters followed by 2 digits) and terminals that fit into those structures. A password's probability is the probability of its structure multiplied by those of its terminals. In 2015 the PCFGs method was integrated with the technique reported by Komanduri in his PhD thesis \cite{komanduri2016modeling}. Conceptually, this method is similar to ours since it also assumes probability independence between model components: Our method divides a given password into $d$ dimensions, and the password's probability is the product of the $d$ probabilities of its corresponding sub-passwords. 
Like the Markov method, the PCFGs method calculates the rank of a given password by enumerating over all possible passwords in descending order of likelihood, so it is also impractical as an online strength estimator. In contrast, our method calculates the rank of a given password in the descending order of likelihood \emph{without enumerating} over the passwords themselves. The PCFGs may be explainable although the authors did not do so.

\cite{li2017personal,houshmandusing,wang2016targeted} extend the PCFGs approach of Weir et al. \cite{weir2009password} to develop systems that also use personal information. The nature of the extension was to add a new grammar variable for each type of personal information, such as B for birthday, N for name and E for email which makes the approach tweakable. These methods are impractical for online use for the same reasons the PCFGs is impractical: the computational limitations of the descending order enumeration.

In 2016 Melicher et al.~\cite{melicher2016fast} proposed to use a recurrent neural network for probabilistic password modeling. Like Markov models, neural networks are trained to generate the next character of a password given the preceding characters of a password. 
In its pure form this method of guess enumeration is similar to that used in Markov models and therefore it is computationally intensive. 

However, the authors also describe a Monte-Carlo method to {\em estimate} the rank of a given password. Therefore, like PESrank, in Monte-Carlo mode the Neural method estimates the rank of a given password without enumerating over the passwords, making it suitable for online strength estimation. The authors do not provide bounds on the estimation error introduced by the Monte Carlo method. However, in order to personalize the neural network method for each user, it should be retrained for each user separately. Since the training takes days, this method cannot be personalized for different users in real-time. Moreover, like most neural-network-based systems, the algorithm is inherently ``un-explainable,'' only providing a numeric rank without any hints about ``why'' or what to do to improve.

In Table~\ref{tbl:relatedWork} we summarize the differences between the leading methods according to several criteria: training time, lookup time, tweaking time, storage, explainabilty and accuracy. The information about PCFGs in this comparison is taken from \cite{weir2009password,komanduri2016modeling} plus \cite{li2017personal,houshmandusing,wang2016targeted} regarding its tweakability.

The table shows that the PESrank method outperforms each of the leading alternatives, in different ways. Unlike PCFGs,
PESrank is online and its training time is significantly shorter. Versus the Neural method in its pure variant again PESrank is superior since it is online, has shorter training time, plus it is tweakable and explainable. Finally, versus the Neural method's Monte-Carlo variant (which is practical as an online estimator), PESrank retains all its other advantages in training and tweak time, explainability, and provable accuracy.

\section{Conclusions} \label{sec:concl}

In this paper we proposed a novel password strength estimator, called PESrank, which accurately models the behavior of a powerful password cracker. PESrank calculates the rank of a given password in an optimal descending sorted order of likelihood. It is practical for {\em online} use, and is able to estimate a given password's rank in fractions of a second. Its training time is drastically shorter than previous methods. Moreover, PESrank is efficiently {\em tweakable} to allow model personalization, without the need to retrain the model; and it is {\em explainable}: it is able to provide information on {\em why} the password has its calculated rank, and gives the user insight on how to pick a better password.

PESrank casts the question of password rank estimation in a multi-dimensional probabilistic framework used in side-channel cryptanalysis, and relies on the ESrank algorithm for side-channel rank estimation. We found that an effective choice uses $d=5$ dimensions: base word, prefix, suffix, capitalization pattern, and l33t transformation. 

We implemented PERrank in Python and and  conducted an extensive evaluation study of it. We also integrated it into the registration page of a course at our university. Even with a model based on 905 million passwords, the response time was well under 1 second, with up to a 1-bit accuracy margin between the upper bound and the lower bound on the rank. 

We conclude that PESrank is a practical strength estimator that can easily be deployed in any web site's online password registration page to assist users in picking better passwords. It provides accurate strength estimates, negligible response-time overhead, good explainability, and reasonable training time.

\bibliographystyle{IEEEtranS}
\bibliography{bibfile.bib}

\end{document}